\documentclass[12pt]{article}
\usepackage{epsfig}
\usepackage{axodraw}
\usepackage{amsmath}
\usepackage{graphics}
\newcommand{\as}{\alpha_{\mathrm{s}}}
\newcommand{\aem}{\alpha_{\mathrm{em}}}
\renewcommand{\d}{\mathrm{d}}

\newcommand{\J}{\mathrm{J}}
\newcommand{\e}{\mathrm{e}}

\newcommand{\g}{\mathrm{g}}
\newcommand{\p}{\mathrm{p}}
\newcommand{\q}{\mathrm{q}}

\newcommand{\qbar}{\mathrm{\overline{q}}}

\newcommand{\kT}{k_{\perp}}
\newcommand{\pT}{p_{\perp}}

\newcommand{\gast}{\gamma^*}
\newcommand{\ga}{\gamma}

\newcommand{\mr}{\mathrm}

\def\Journal#1#2#3#4{{#1}{\bf #2} (#3) #4}

\def\NPB{{\it Nucl. Phys.~}{\bf B}}
\def\PLB{{\it Phys. Lett.~}{\bf  B}}
\def\JournalPLB#1#2#3{{\it Phys. Lett.~}{\bf {#1}B} (#2) #3}

\def\ZPC{{\it Z. Phys.~}{\bf C}}

\def\JHEP{\it J. High Energy Phys.~}

\def\CPC{\it Computer Phys. Commun.~}
\def\PRP{\it Phys. Rept.~}

\def\EPJC{{\it Eur. Phys. J.~}{\bf C}}

\def\SJNP{\it Sov. J. Nucl. Phys.~}
\def\SPJP{\it Sov. Phys. JETP~}

\newlength{\abstwidth}
\newlength{\captivewidth}
\newcommand{\captive}[1]{\rule{5mm}{0mm}%
\begin{minipage}{\captivewidth}%
\caption[small]{#1}\end{minipage}}
\begin{document} 
%

\sloppy
 
\pagestyle{empty}

\begin{flushright}
LU TP 00--45\\
hep-ph/0010264\\
Oktober 2000
\end{flushright}
 
\vspace{\fill}
 
\begin{center}
{\LARGE\bf Total Cross Sections and Event Properties\\[2mm]
from Real to Virtual Photons$^\star$}\\[10mm]
{\Large 
Christer Friberg
}\\[2mm]
{\it Department of Theoretical Physics,}\\[1mm]
{\it Lund University, Lund, Sweden}\\[1mm]
christer@thep.lu.se
\end{center}

\vspace{\fill}
\begin{center}
{\bf Abstract}\\[2ex]
\begin{minipage}{\abstwidth}
A model for total cross sections with virtual photons is presented. 
In particular $\gast\p$ and $\gast\gast$ cross sections are considered. 
Our approach extends on a model for photoproduction, where the total 
cross section is subdivided into three distinct event classes: 
direct, VMD and anomalous. 
With increasing photon virtuality, the latter two decrease in importance.
Instead Deep Inelastic Scattering dominates, with the direct class being 
the $\mathcal{O}(\alpha_{\mr{s}})$ correction thereof. 
Hence, the model provides a smooth transition between the 
two regions. By the breakdown into different event classes, one may
aim for a complete picture of all event properties.
\end{minipage}
\end{center}

\vspace{\fill}

\footnoterule
{\footnotesize $^\star$To appear in the Proceedings of the %
International Conference on %
the Structure and Interactions %
of the Photon; Photon 2000, 26th-31st August 2000, %
Ambleside, England.}

\clearpage
\pagestyle{plain}
\setcounter{page}{1}

\section*{Introduction}

In this section we summarize the model presented in \cite{ourjet,ourtot}.
It starts from the model for real photons in \cite{SaSmodel}, but further
develops this model and extends it also to encompass the physics of
virtual photons. The physics has been implemented in the \textsc{Pythia}
generator \cite{pythia}, so that complete events can be studied under 
realistic conditions.   

Photon interactions are complicated since the photon wave function
contains so many components, each with its own interactions. To
first approximation, it may be subdivided into a direct and a resolved
part. (In higher orders, the two parts can mix, so one has 
to provide sensible physical separations between the two.)
In the former the photon acts as a pointlike particle, 
while in the latter it fluctuates into hadronic states.
These fluctuations are of $\mathcal{O}(\alpha_{\mathrm{em}})$, and so
correspond to a small fraction of the photon wave function, but this
is compensated by the bigger cross sections allowed in strong-interaction
processes. For real photons therefore the resolved processes dominate
the total cross section, while the pointlike ones take over for 
virtual photons. 
 
\section*{A Model for Photon Interactions}

The fluctuations $\gamma \to \q\qbar \, (\to \gamma)$ can be characterized 
by the transverse momentum $\kT$ of the quarks, or alternatively by some
mass scale $m \simeq 2 \kT$, with a spectrum of fluctuations 
$\propto \d\kT^2/\kT^2$. The low-$\kT$ part cannot be calculated 
perturbatively, but is instead parameterized by experimentally determined  
couplings to the lowest-lying vector mesons, $V = \rho^0$, $\omega^0$, 
$\phi^0$ and $\J/\psi$, an ansatz called VMD for Vector Meson 
Dominance. Parton distributions are defined with a unit
momentum sum rule within a fluctuation \cite{SaSpdf}, giving rise
to total hadronic cross sections, jet activity, multiple interactions 
and beam remnants as in hadronic interactions. In interactions with a hadron 
or another resolved photon, jet production occurs by typical 
parton-scattering processes such as $\q\q' \to \q\q'$ or $\g\g \to \g\g$.

States at larger $\kT$
are called GVMD or Generalized VMD, and their contributions to the 
parton distribution of the photon are called anomalous. Given a dividing 
line $k_0 \simeq 0.5$~GeV to VMD states, the anomalous parton distributions 
are perturbatively calculable. The total cross section of a state is not, 
however, since this involves aspects of soft physics and eikonalization 
of jet rates. Therefore an ansatz is chosen where the total cross section 
of a state scales like $k_V^2/\kT^2$, where the adjustable parameter 
$k_V \approx m_{\rho}/2$ for light quarks. The spectrum of GVMD states is taken 
to extend over a range $k_0 < \kT < k_1$, where $k_1$ is identified with 
the $p_{\perp\mathrm{min}}(s)$ cut-off of the perturbative jet spectrum in 
hadronic interactions, $p_{\perp\mathrm{min}}(s) \approx 1.5$~GeV at typical 
energies \cite{pythia}. Above that range, the states are assumed to be 
sufficiently weakly interacting that no eikonalization procedure is required,
so that cross sections can be calculated perturbatively without any recourse
to Pomeron phenomenology. There is some arbitrariness in that choice, and 
some simplifications are required in order to obtain a manageable description.
 
A real direct photon in a $\gamma\p$ collision can interact with the parton
content of the proton: $\gamma\q \to \q\g$ (QCD Compton) and 
$\gamma\g \to \q\qbar$ (Boson Gluon Fusion). The $\pT$ in this collision is 
taken to exceed $k_1$, in order to avoid double-counting with the interactions 
of the GVMD states. In $\gamma\gamma$, the equivalent situation is called 
single-resolved, where a direct photon interacts with the partonic component
of the other, resolved photon. The $\gamma\gamma$ direct process
$\gamma\gamma \to \q\qbar$ has no correspondence in $\gamma\p$.

\begin{figure}[t]
\begin{center}
\resizebox{160mm}{!}{
\begin{picture}(105,200)(-4,-25)
  \Photon(7,140)(45,120){4}{4}  
  \GOval(12,10)(10,5)(0){0.5}
  \DashLine(17,5)(90,5){4}
  \ArrowLine(17,13)(45,30)
  \Gluon(45,75)(45,30){4}{4}
  \ArrowLine(45,75)(45,120)
  \ArrowLine(45,30)(90,30)
  \ArrowLine(90,75)(45,75)
  \ArrowLine(45,120)(90,120) 
  \Text(0,10)[]{$\p$}
  \Text(0,140)[]{$\ga$}     
  \Text(60,50)[]{$\pT$} 
  \Text(60,95)[]{$\kT$} 
  \Text(97,30)[]{$\q'$} 
  \Text(97,75)[]{$\qbar$} 
  \Text(97,120)[]{$\q$} 
\end{picture}  
\begin{picture}(200,200)(0,0)
  \LongArrow(15,15)(185,15)
  \Text(195,15)[]{$\kT$} 
  \LongArrow(15,15)(15,185) 
  \Text(15,194)[]{$\pT$} 
  \SetWidth{1.5}
  \Line(50,15)(50,180)
  \Text(50,5)[]{$k_0$}
  \Line(100,15)(100,100)
  \Text(100,5)[]{$k_1$}
  \Line(100,100)(180,180)
  \Text(180,190)[]{$\kT = \pT$}
  \Text(32,90)[]{VMD}
  \Text(75,100)[]{GVMD}
  \Text(140,60)[]{direct}
\end{picture}   
\hspace{3ex}
\begin{picture}(105,220)(-4,-15)
  \ArrowLine(5,165)(45,165)
  \ArrowLine(45,165)(90,180)
  \Photon(45,165)(45,120){4}{4}  
  \GOval(12,10)(10,5)(0){0.5}
  \DashLine(17,5)(90,5){4}
  \ArrowLine(17,13)(45,30)
  \Gluon(45,75)(45,30){4}{4}
  \ArrowLine(45,75)(45,120)
  \ArrowLine(45,30)(90,30)
  \ArrowLine(90,75)(45,75)
  \ArrowLine(45,120)(90,120) 
  \Text(0,10)[]{$\p$}
  \Text(0,165)[]{$\e$}
  \Text(30,140)[]{$\ga^*$}     
  \Text(30,50)[]{$\g$}     
  \Text(60,50)[]{$\pT$} 
  \Text(60,95)[]{$\kT$} 
  \Text(60,140)[]{$Q$} 
  \Text(97,30)[]{$\q'$} 
  \Text(97,75)[]{$\qbar$} 
  \Text(97,120)[]{$\q$} 
  \Text(97,180)[]{$\e'$} 
\end{picture}   
\begin{picture}(200,200)(0,0)
  \LongArrow(15,15)(185,15)
  \Text(195,15)[]{$\kT$} 
  \LongArrow(15,15)(15,185) 
  \Text(15,194)[]{$\pT$} 
  \SetWidth{1.5}
  \Line(15,15)(180,180)
  \Line(100,15)(100,180)
  \Text(100,5)[]{$Q$}
  \Text(180,190)[]{$\kT = \pT$}
  \Text(70,35)[]{LO DIS}
  \Text(150,60)[]{direct}
  \Text(55,100)[]{non-DGLAP}
  \Text(135,170)[]{resolved $\gast$}
\end{picture} 
}  
\end{center}
{\small
\vspace{-2mm}
\hspace{1cm}(a)\hspace{3cm}(b)\hspace{4.5cm}(c)\hspace{3cm}(d)
}
\vspace{2mm}
\captive%
{(a) Schematic graph for a hard $\ga\p$ process, illustrating
the concept of two different scales. 
(b) The allowed phase space for this process, with one subdivision
into event classes.
(c) Schematic graph for a hard $\ga^*\p$ process, illustrating
the concept of three different scales. 
(d) Event classification in the large-$Q^2$ limit.
\label{fig:gammapplane}
\label{fig:gastpplane}
}
\end{figure}

As an illustration of this scenario, the phase space of $\gamma\p$ events is 
shown in Fig.~\ref{fig:gammapplane}a-b. (A corresponding plot can be made for 
$\gamma\gamma$, but then requires three dimensions.)
Two transverse momentum scales are introduced, namely the 
photon resolution scale $\kT$ and the hard interaction scale $\pT$.
Here $\kT$ is a measure of the virtuality of a fluctuation of the photon 
and $\pT$ corresponds to the most virtual rung of the ladder, 
possibly apart from $\kT$. 
As we have discussed above, the low-$\kT$ region corresponds to
VMD and GVMD states that encompasses both perturbative high-$\pT$ and
non-perturbative low-$\pT$ interactions. Above $k_1$, the region is split 
along the line $\kT = \pT$. When $\pT > \kT$ the photon is resolved by
the hard interaction, as described by the anomalous part of the photon 
distribution function. This is as in the GVMD sector, except that we should 
(probably) not worry about multiple parton--parton interactions. In the
complementary region $\kT > \pT$, the $\pT$ scale is just part of the 
traditional evolution of the proton PDF's up to the scale of $\kT$, and thus
there is no need to introduce an internal structure of the photon. 
One could imagine the direct class of events as extending below $k_1$
and there being the low-$\pT$ part of the GVMD class, only appearing 
when a hard interaction at a larger $\pT$ scale would not preempt it.  
This possibility is implicit in the standard cross section framework.  

If the photon is virtual, it has a reduced probability to fluctuate into 
a vector meson state, and this state has a reduced interaction probability.
This can be modeled by a traditional dipole factor
$(m_V^2/(m_V^2 + Q^2))^2$ for a photon of virtuality $Q^2$, where 
$m_V \to 2 \kT$ for a GVMD state. Putting it all together, the cross
section of the GVMD sector then scales like
\begin{equation}
\int_{k_0^2}^{k_1^2} \frac{\d\kT^2}{\kT^2} \, \frac{k_V^2}{\kT^2} \,
\left( \frac{4\kT^2}{4\kT^2 + Q^2} \right)^2 ~.
\end{equation}

For a virtual photon the DIS process $\gast \q \to \q$
is also possible, but by gauge invariance its cross section must
vanish in the limit $Q^2 \to 0$. At large $Q^2$, the direct processes 
can be considered as the $\mathcal{O}(\as)$ correction to the lowest-order 
DIS process, but the direct ones survive for $Q^2 \to 0$. There is no 
unique prescription for a proper combination at all $Q^2$, but we have
attempted an approach that gives the proper limits and minimizes 
doublecounting. For large $Q^2$, the DIS $\gast\p$ cross section
is proportional to the structure function $F_2 (x, Q^2)$ with the
Bjorken $x = Q^2/(Q^2 + W^2)$. Since normal parton distribution 
parameterizations are frozen below some $Q_0$ scale and therefore do not
obey the gauge invariance condition, an ad hoc factor 
$(Q^2/(Q^2 + m_{\rho}^2))^2$ is introduced for the conversion from 
the parameterized $F_2(x,Q^2)$ to a $\sigma_{\mr{DIS}}^{\gast\p}$:
\begin{equation}
\sigma_{\mathrm{DIS}}^{\gast\p} \simeq  
\left( \frac{Q^2}{Q^2 + m_{\rho}^2} \right)^2 \,
\frac{4\pi^2\aem}{Q^2} F_2(x,Q^2) =
\frac{4\pi^2\aem Q^2}{(Q^2+m_\rho^2)^2} \,
\sum_{\q ,\qbar} e_{\q}^2 \, 
x  q(x, Q^2) 
~.
\label{sigDIS}
\end{equation}
Here $m_\rho$ is some non-perturbative hadronic mass parameter, for 
simplicity identified with the $\rho$ mass. 

In order to avoid double-counting between DIS and direct events, a requirement 
$\pT > \max(k_1, Q)$ is imposed on direct events. In the remaining DIS ones, 
denoted lowest order (LO) DIS, thus $\pT < Q$. This would suggest a subdivision
$\sigma_{\mr{LO\,DIS}}^{\gast\p} = \sigma_{\mr{DIS}}^{\gast\p} -
\sigma_{\mr{direct}}^{\gast\p}$, with $\sigma_{\mr{DIS}}^{\gast\p}$ given by
eq.~(\ref{sigDIS}) and $\sigma_{\mr{direct}}^{\gast\p}$ by the perturbative
matrix elements. In the limit $Q^2 \to 0$, the DIS cross section is now
constructed to vanish while the direct is not, so this would suggest
$\sigma_{\mr{LO\,DIS}}^{\gast\p} < 0$. However, here we expect the correct 
answer not to be a negative number but an exponentially suppressed one, 
by a Sudakov form factor. This modifies the cross section: 
\begin{equation}
\sigma_{\mr{LO\,DIS}}^{\gast\p} = \sigma_{\mr{DIS}}^{\gast\p} -
\sigma_{\mr{direct}}^{\gast\p}
~~ \longrightarrow ~~ 
\sigma_{\mr{DIS}}^{\gast\p} \; \exp \left( - \frac{%
\sigma_{\mr{direct}}^{\gast\p}}{\sigma_{\mr{DIS}}^{\gast\p}} \right) \;.
\label{eq:LODIS}
\end{equation}
Since we here are in a region where the DIS cross section is no longer the 
dominant one, this change of the total DIS cross section is not essential. 

The overall picture, from a DIS perspective, is illustrated in 
Fig.~\ref{fig:gastpplane}c-d,
now with three scales to be kept track of. The traditional DIS region 
is the strongly ordered one, $Q^2 \gg \kT^2 \gg \pT^2$, where 
DGLAP-style evolution \cite{DGLAP} is responsible for the event 
structure. As always, ideology wants strong ordering, while 
the actual classification is based on ordinary ordering 
$Q^2 > \kT^2 > \pT^2$. The region $\kT^2 > \max(Q^2,\pT^2)$ is also
DIS, but of the $\mathcal{O}(\as)$ direct kind. The region 
where $\kT$ is the smallest scale corresponds to 
non-ordered emissions, that then go beyond DGLAP validity,
while the region $\pT^2 > \kT^2 > Q^2$ cover the interactions of a 
resolved virtual photon. Comparing Figs.~\ref{fig:gammapplane}b
and \ref{fig:gastpplane}d, we conclude that the whole region
$\pT > \kT$ involves no doublecounting, since we have made no
attempt at a non-DGLAP DIS description but can choose to cover this 
region entirely by the VMD/GVMD descriptions. Actually, it is only 
in the corner $\pT < \kT < \min(k_1, Q)$ that an overlap can occur 
between the resolved 
and the DIS descriptions. Some further considerations show that
usually either of the two is strongly suppressed in this region,
except in the range of intermediate $Q^2$ and rather small $W^2$.
Typically, this is the region where $x \approx Q^2/(Q^2 + W^2)$ is not 
close to zero, and where $F_2$ is dominated by the valence-quark 
contribution. The latter behaves roughly $\propto (1-x)^n$, with an 
$n$ of the order of 3 or 4. Therefore we will introduce a corresponding 
damping factor to the VMD/GVMD terms. 

In total, we have now arrived at our ansatz for all $Q^2$:
\begin{equation}
\sigma_{\mr{tot}}^{\gast\p} = 
\sigma_{\mr{DIS}}^{\gast\p} \; \exp \left( - 
\frac{\sigma_{\mr{direct}}^{\gast\p}}{\sigma_{\mr{DIS}}^{\gast\p}} \right) +
\sigma_{\mr{direct}}^{\gast\p} +
\left( \frac{W^2}{Q^2 + W^2} \right)^n \left(
\sigma_{\mr{VMD}}^{\gast\p} + 
\sigma_{\mr{GVMD}}^{\gast\p} \right) \;,
\end{equation}
with four main components. Most of these in their turn have
a complicated internal structure, as we have seen. The $\gast\gast$ 
collision between two inequivalent photons contains 13 components: four 
when the VMD and GVMD states interact with each other (double-resolved),
eight with a LO~DIS or direct photon interaction on a VMD or GVMD state on
either side (single-resolved, including the traditional DIS), and one 
where two direct photons interact by the  process $\gast\gast \to \q\qbar$ 
(direct, not to be confused with the direct process of $\gast\p$). 

An important note is that the $Q^2$ dependence of the DIS and direct 
processes is implemented in the matrix element expressions, i.e. in 
processes such as $\gast\gast \to \q\qbar$ or $\gast \q \to \q\g$
the photon virtuality explicitly enters. This is different from 
VMD/GVMD, where dipole factors are used to reduce the assumed flux of 
partons inside a virtual photon relative to those of a real one, but 
the matrix elements themselves contain no dependence on the virtuality
either of the partons or of the photon itself. Typically results are 
obtained with the SaS~1D PDF's for the virtual (transverse) photons
\cite{SaSpdf}, since these are well matched to our framework, e.g.
allowing a separation of the VMD and GVMD/anomalous components. 

%
\section*{Results}

So far, nothing has been said about $\sigma_{\mr{L}}$, except that gauge
invariance dictates its vanishing in the limit $Q^2 \to 0$. However, we will
assume that quite a similar decomposition can be made of longitudinal photon
interactions as was done for the transverse one. To first approximation, 
this again means a separation into direct and resolved photons.
In direct processes, the nature of the photon is explicitly included in the 
perturbative cross section formulae. Thus, for $\gast\q \to \q\g$ and
$\gast \g\to\q\qbar$, the differential cross sections  
$\d\hat{\sigma}_{\mathrm{T}} / \d\hat{t}$ and
$\d\hat{\sigma}_{\mathrm{L}} / \d\hat{t}$ are separately available 
\cite{siggap}. The latter is proportional to $Q^2$ and thus nicely
vanishes in the limit $Q^2 \to 0$. Similarly the $\gast\gast \to \q\qbar$
process gives four separate cross section formulae,
$\d\hat{\sigma}_{\mathrm{TT,TL,LT,LL}} / \d\hat{t}$ \cite{siggaga}.
The DIS, non-direct part currently contains no explicit description of a
longitudinal probing photon, only of a probed one.
However, to the extent that PDF's are extracted from 
$F_2 \propto \sigma_{\mathrm{T}} + \sigma_{\mathrm{L}}$ data,
effects may be implicitly included. Furthermore, perturbative calculations
\cite{RinDISthy} predict $\sigma_{\mathrm{L}} \ll \sigma_{\mathrm{T}}$
in the large-$Q^2$ region, where this process dominates. 

\begin{figure} 
\mbox{
\hspace{-0.5cm}\epsfig{file=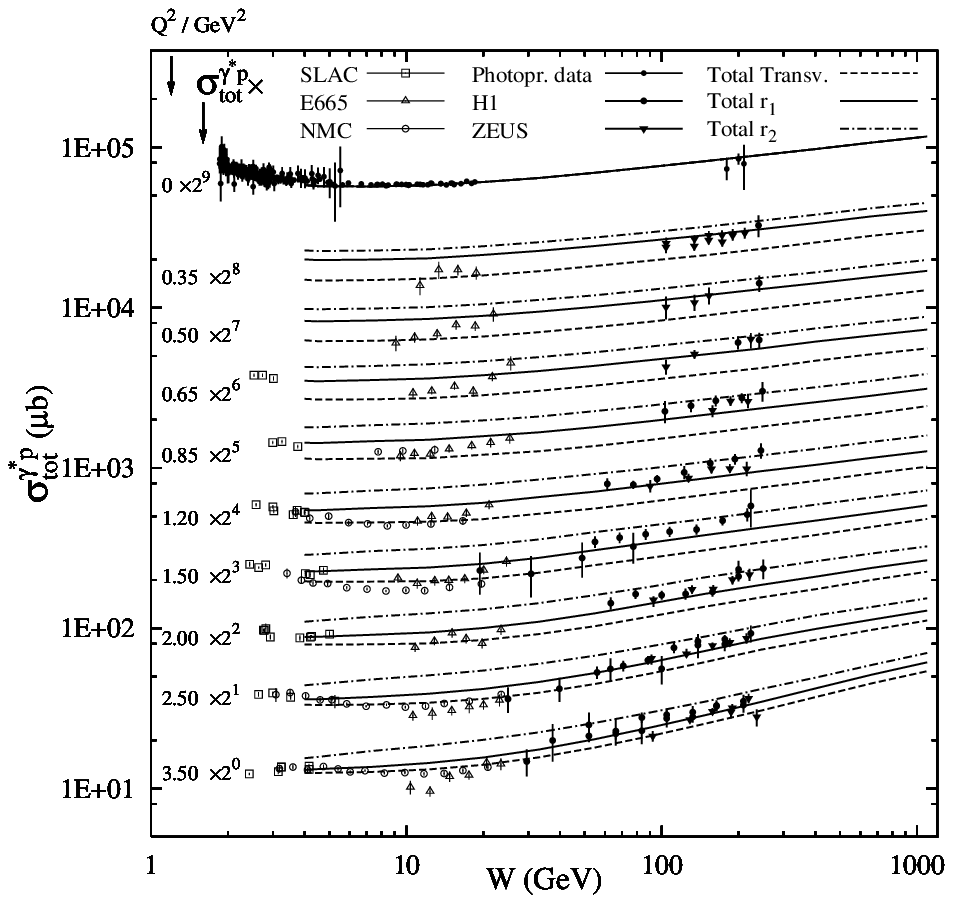,width=10.9cm}
\hspace{-2.7cm}\epsfig{file=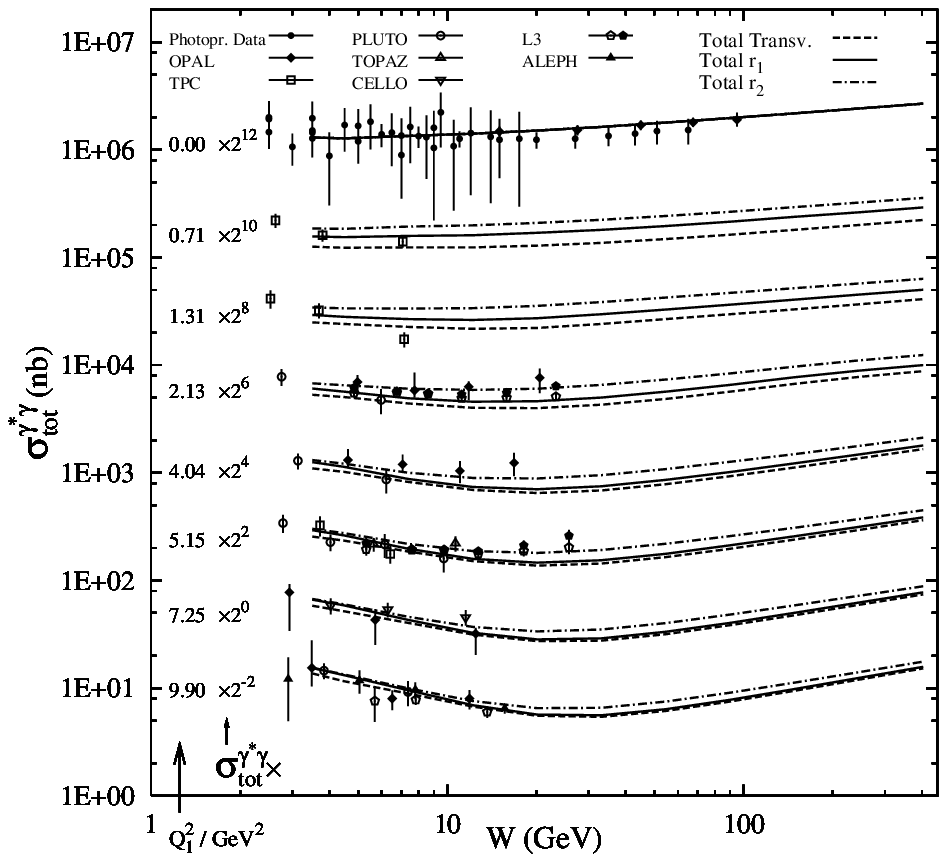,width=10.9cm}}
\captive{Total cross sections as a function of the invariant mass of the 
collision. References to the experimental measurements can be found in
ref.~\protect\cite{ourtot}.
\label{fig:Wdep}
}
\end{figure}

In \cite{ourjet,ourtot,ourgL} a few simple $Q^2$-dependent multiplicative 
expressions have been studied to encompass the effects of resolved 
longitudinal photons. The two alternative factors 
\begin{equation}
r_1(m_V^2, Q^2) = \frac{2 m_V^2 Q^2}{(m_V^2 + Q^2)^2} \qquad
r_2(m_V^2, Q^2) = \frac{2 Q^2}{(m_V^2 + Q^2)}
\label{eq:R}
\end{equation}
relative to the resolved transverse cross section
are used for the resolved longitudinal contributions in Fig.~\ref{fig:Wdep}, 
assuming a relation $m_V \simeq 2 \kT$ for GVMD states. The differences
between the $r_i$ factors show some of the uncertainty in the modeling 
of resolved longitudinal photons. However, it is possible to obtain a 
reasonable description of all the data in both $\gast\p$ and $\gast\gamma$
with the same set of parameters (mainly constrained from $\gamma\p$).

More sophisticated tests of the model can be made by studying event shapes and 
we look forward to detailed studies by the experimental community, based on 
the code we now provide in the {\sc Pythia} event generator~\cite{pythia}. 

\newpage

\begin{thebibliography}{99}

\bibitem{ourjet}
C. Friberg and T. Sj\"ostrand,
\Journal{\EPJC}{13}{2000}{151}
(hep-ph/9907245).

\bibitem{ourtot}
C. Friberg and T. Sj\"ostrand,
\Journal{\JHEP}{09}{2000}{010} 
(hep-ph/0007314).

\bibitem{SaSmodel}
G.A. Schuler and T. Sj\"ostrand, 
\Journal{\PLB}{300}{1993}{169}, \\
\Journal{\NPB}{407}{1993}{539},
\Journal{\ZPC}{73}{1997}{677}.

\bibitem{pythia} 
T. Sj\"ostrand, 
\Journal{\CPC}{82}{1994}{74};\\
T. Sj\"ostrand et al., 
hep-ph/0010017 (to appear in {\CPC});\\
http://www.thep.lu.se/$\sim\,$torbjorn/Pythia.html.

\bibitem{SaSpdf}
G.A. Schuler and T. Sj\"ostrand, 
\Journal{\ZPC}{68}{1995}{607}, \\
\Journal{\PLB}{376}{1996}{193}.

\bibitem{DGLAP}
V.N. Gribov and L.N. Lipatov, 
\Journal{\SJNP}{15}{1972}{438 and 675};\\
G.~Altarelli and G.~Parisi, 
\Journal{\NPB}{126}{1977}{298};\\
Yu.L.~Dokshitzer, 
\Journal{\SPJP}{46}{1977}{641}.

\bibitem{siggap}
G. Altarelli and G. Martinelli, 
\JournalPLB{76}{1978}{89};\\
A. Mend\'ez, 
\Journal{\NPB}{145}{1978}{199};\\
R. Peccei and R. R\"uckl, 
\Journal{\NPB}{162}{1980}{125};\\
Ch. Rumpf, G. Kramer and J. Willrodt, 
\Journal{\ZPC}{7}{1981}{337}.

\bibitem{siggaga}
V.M.~Budnev, I.F.~Ginzburg, G.V.~Meledin and V.G.~Serbo, \\ 
\Journal{\PRP}{15}{1974}{181};\\
V.N.~Baier, E.A.~Kuraev, V.S.~Fadin and V.A.~Khoze, 
\Journal{\PRP}{78}{1981}{293}.

\bibitem{RinDISthy}
S.~R.~Mishra and F.~Sciulli,
\Journal{\PLB}{244}{1990}{341}.

\bibitem{ourgL}
C. Friberg and T. Sj\"ostrand,
LU TP 00--31, to appear in {\PLB} \\
(hep-ph/0009003).

\end{thebibliography}
\end{document}